\documentclass[conf]{new-aiaa}
\usepackage[utf8]{inputenc}

\usepackage{graphicx}
\usepackage{amsmath}
\usepackage[version=4]{mhchem}
\usepackage{siunitx}
\usepackage{longtable,tabularx}
\usepackage{caption}
\usepackage{subcaption}
\usepackage{wrapfig}
\usepackage{float}

\title{Fully Compressible Magnetohydrodynamic Simulations of Solar Convection Zones with CHORUS++ 
}

\author{Aidan Paoli\footnote{Undergraduate Student, AIAA Student Member, Membership Number: 1400547} and Chunlei Liang \footnote{Professor, AIAA Associate Fellow, Membership Number: 303971}}
\affil{Department Mechanical and Aerospace Engineering, Clarkson University, Potsdam, NY, United States}

\begin{document}

\maketitle

\begin{abstract}
The objective of this study is to develop a fully compressible magnetohydrodynamic solver for fast simulations of the global dynamo of the Sun using unstructured grids and GPUs. Solar activity largely dictates Earth's immediate space environment and atmosphere. Accurate modeling of the Sun's convective layers is vital to predicting the Sun's behavior, including solar dynamo and sunspot cycles. Similarly, better understanding of convection in gas giants like Jupiter and Saturn requires global simulations of their convective layers. Global simulations modeling convective layers inside stars and planets are computationally strenuous. Currently, there are many efficient codes capable of conducting these large simulations; however, many make assumptions of anealastic density distribution. The anelastic assumption is capable of producing accurate results for low mach numbers; however, it fails in regions with a higher mach number and a fully compressible flow must be considered. Many of these codes are also required to use a structured grid for the use of spherical harmonics. To avoid these issues, Wang et al. \cite{Wang2015} created a Compressible High-ORder Unstructured Spectral difference (CHORUS) code. CHORUS is a fully compressible, high-order, spectral difference code for simulating fluid dynamics inside stars and planets. CHORUS++ augmented the CHORUS code to adopt a higher degree of polynomials by using cubed-sphere meshing and transfinite mapping to perform simulations on unstructured grids \cite{Chen2023}. Two hydrodynamic benchmark tests of Jupiter and the Sun were used to test the CHORUS code \cite{Wang2015,Chen2023}. Recently, CHORUS++ was further developed for parallel magnetohydrodynamic (MHD) solutions on GPUs at Clarkson University. This study presents CHORUS-MHD solutions for two dynamo benchmark problems similar to the one proposed by Chen et al. \cite{Chen2023}. We extended the solar benchmark problems presented by Chen et al. \cite{Chen2023} to unsteady solar dynamo problems, with two different density scale heights. The CHORUS-MHD code is further accelerated by GPUs and used to successfully solve these solar dynamo benchmark problems. Both solar benchmark problems are run using a 6th-order spectral difference method on multiple GPUs. CHORUS-MHD can be further used to simulate the sunspot cycle and model the polar vortices of the Sun \cite{Dikpati2024} because of its flexibility in meshing.

\end{abstract}

\section{Introduction}
The formation and development of the solar system were largely dictated by the Sun. Today, the immediate space environment is heavily influenced by solar activity, which is closely correlated to magnetic field activity in the Sun, sunspots, and other solar phenomena. By improving our understanding of the Sun, we can ultimately begin to predict its behavior and understand the evolution of the solar cycle including sunspot cycles, solar flares, and magnetic field polarity. To better understand these phenomena, we must investigate the convective layers of the Sun, where large amounts of charged plasma is moved, producing magnetic fields. Modeling these convective layers is often computationally taxing and difficult due to the vast difference in length scales and timescales in planets and stars, though there are multiple successful approaches. Many models use an anelastic density approximation, allowing for the implementation of spherical harmonics to solve governing equations; however, there are drawbacks. First, the anelastic approximation and spherical harmonics require the code to be used on spherically structured shells, which results in lower meshing quality at the poles. Simultaneously, the anelastic approximation begins to lose accuracy as the Mach number approaches one, as seen in convection in red giants and modeling of convection into the photosphere where sunspots and granulation occurs. To accurately model the convection of the Sun with a fully compressible code, CHORUS was developed. CHORUS is a fully Compressible High-ORder Unstructured Spectral difference code for modeling fluid dynamics in spherical shells, originally developed by Wang et al. \cite{Wang2015}. CHORUS uses conservation of mass, energy, and momentum equations to solve hydrodynamic problems in spherical shells. CHORUS++ extended the code to include transfinite mapping to a computational domain, and an unstructured cubed sphere meshing technique. The unstructured nature allows CHORUS to work for oblate shells such as Saturn and other rapidly rotating bodies. The cubed-sphere meshing technique also provides a more uniform and refined grid resolution, especially in the polar regions, and singularities can be avoided at the north and south poles. In smaller-size stars, where deep convection is typically at a low Mach number and incompressible flow can be assumed, the fully compressible nature of CHORUS is not ideal. However, CHORUS can simulate problems that are not achievable by incompressible codes such as red giant convection, photospheric convection with sunspots, and granulation, where compressible flow needs to be considered. Recently, CHORUS has been extended to CHORUS-MHD, which includes additional induction equations and an equation to clean the divergence of the magnetic field \cite{Chen2023b}. CHORUS-MHD is also capable of being run on multiple GPUs to speed up simulation time.

\section{MHD Governing Equations}
CHORUS considers the fluid dynamics in a hollow spherical shell. The reference frame of the simulation rotates uniformly about the z-axis with the spherical shell's angular speed. It is assumed that most of the mass is concentrated inside the shell, resulting in simpler gravitational terms. The governing equations consist of conservation equations for mass, total energy, linear momentum, and the magnetic induction, in addition to a divergence cleaning equation of the magnetic field. The general form of these equations can be expressed in Equation \ref{divergence form}.

\begin{equation}
    \frac{\partial \mathbf{Q}}{\partial t} + \mathbf{\nabla} \cdot \overline{\mathbf{F}} = \mathbf{M}
    \label{divergence form}
\end{equation}

Where $\mathbf{Q}$ is the vector of conserved variables, $\mathbf{M}$ is the vector of the source term, and $\mathbf{F}$ is the flux vector with components $F\hat{\mathbf{x}}$, $G\hat{\mathbf{y}}$, $H\hat{\mathbf{z}}$. $\mathbf{Q}$ is defined in Equation \ref{General Form1}.
\begin{equation}
    Q = \begin{pmatrix} \rho \\ \rho \mathbf{U} \\ E \\ \mathbf{B} \\ \psi \end{pmatrix}
    \label{General Form1}
\end{equation}

Where $\rho$ is the density, $\mathbf{U}$ is the velocity vector, $E$ is the total energy, $\mathbf{B}$ is the magnetic field vector and $\mathbf{\psi}$ is the divergence cleaning term of the generalized Lagrange multiplier (GLM) for the magnetic field \cite{Derigs2018}. The total energy density is composed of internal energy, kinetic energy, magnetic field energy, and a divergence cleaning term. The total energy density is defined as
\begin{equation}
    E = e + \frac{1}{2} \psi^2 = \frac{p}{\gamma - 1} + \frac{1}{2} \rho \| \mathbf{U} \|^2 + \frac{1}{2} \| \mathbf{B} \|^2 + \frac{1}{2} \psi^2
    \label{eq:energy_equation}
\end{equation}
Where $p$ is the pressure, $\gamma$ is the ratio of specific heats, and $\| \cdot \|$ is the Euclidean vector norm. We assume an ideal gas so that 
\begin{equation}
    p = \rho RT
    \label{ideal gas}
\end{equation}
With $R$ being the specific gas constant, and $T$ is the temperature. The vector of sources term consists of terms for the gravitational force and the gravitational potential energy, Coriolis force, and additional terms for cleaning divergence \cite{Derigs2018}. The complete source vector is 
\begin{equation}
M = \begin{pmatrix}
   0 \\
   \rho \mathbf{g} - 2 \rho \mathbf{\Omega} \times \mathbf{U} - (\mathbf{\nabla} \cdot \mathbf{B}) \mathbf{B} \\
   \rho \mathbf{U} \cdot \mathbf{g} - (\mathbf{\nabla} \cdot \mathbf{B}) \mathbf{U} \cdot \mathbf{B} \\
   -(\mathbf{\nabla} \cdot \mathbf{B}) \mathbf{U} \\
   -a \psi 
\end{pmatrix}
\end{equation}
Where $\mathbf{g}=-\frac{GM}{r^2} \mathbf{\hat{r}}$ is the acceleration due to gravity, and $\mathbf{\Omega}=\Omega\mathbf{\hat{z}}$ is the rotation rate. The flux term $\mathbf{\bar{F}}$ is composed of inviscid and viscous fluxes as in $\mathbf{\bar{F}}=\mathbf{\bar{F}}_{inv}-\mathbf{\bar{F}}_{vis}$. Inviscid and viscous fluxes are shown in Equation \ref{invF} and Equation \ref{visF}.
\begin{equation}
    \mathbf{\bar{F}_{inv}} = 
    \begin{pmatrix}\rho \mathbf{U} \\
        \rho \mathbf{U} \otimes \mathbf{U} + \left( p + \frac{1}{2} \|\mathbf{B}\|^2 \right) I - \mathbf{B} \otimes \mathbf{B} \\
        \mathbf{U} \left( e + p + \frac{1}{2} \|\mathbf{B}\|^2 \right) - \mathbf{B} (\mathbf{U} \cdot \mathbf{B}) + C_h \mathbf{\psi} \\
        \mathbf{U} \otimes \mathbf{B} - \mathbf{B} \otimes \mathbf{U} + C_h \mathbf{\mathbf{\psi}} I \\
        C_h \mathbf{B}
    \end{pmatrix} \\
    \label{invF}
\end{equation}

\begin{equation}
    \mathbf{\bar{F}_{vis}} = 
    \begin{pmatrix}
        0 \\
        \bar{\mathbf{\mathbf{\tau}}} \\
        \mathbf{U} \cdot \bar{\mathbf{\tau}} - \mathbf{q} + \eta (\mathbf{B} \times \mathbf{J}) \\
        \eta (\mathbf{\nabla} \times \mathbf{B}) \\
        0
    \end{pmatrix}
    \label{visF}
\end{equation}
In Equations \ref{invF} and \ref{visF}, $I$ is the identity matrix, $\eta$ is the magnetic diffusivity, $\mathbf{J}$ is the current density defined as $\mathbf{J}=\mathbf{\nabla}\times\mathbf{B}$, $\mathbf{\bar\tau}$ is the shear stress tensor, and $\mathbf{q}$ is the heat flux. The shear stress tensor and heat flux are defined in Equations \ref{ssTensor} and \ref{qFlux} respectively.
\begin{equation}
        \mathbf{\overline{\tau}} = \mu (\mathbf{\nabla} \mathbf{U} + (\mathbf{\nabla} \mathbf{U})^T) + \lambda (\mathbf{\nabla} \cdot \mathbf{U}) I
        \label{ssTensor}
\end{equation}
\begin{equation}
        \mathbf{q} = -\kappa \rho T \mathbf{\mathbf{\nabla}} S - \kappa_r \rho C_p \mathbf{\nabla} T
        \label{qFlux}
\end{equation}
$\mu$ is the dynamic viscosity, $\lambda= -\frac{2}{3} \mu$ based on Stokes' hypothesis, $\kappa$ is the thermal diffusivity, $\kappa_{r}$ is the radiative diffusivity, and $S$ is the specific entropy. The specific entropy is given in Equation \ref{eq:entropy}.

\begin{equation}
  S = C_p \ln \left( \frac{p^\frac{1}{\gamma}}{\rho} \right)
  \label{eq:entropy}
\end{equation}

\section{Meshing}
\begin{figure}[h]
     \centering
     \begin{subfigure}[b]{0.48\textwidth}
         \centering
         \includegraphics[width=\textwidth]{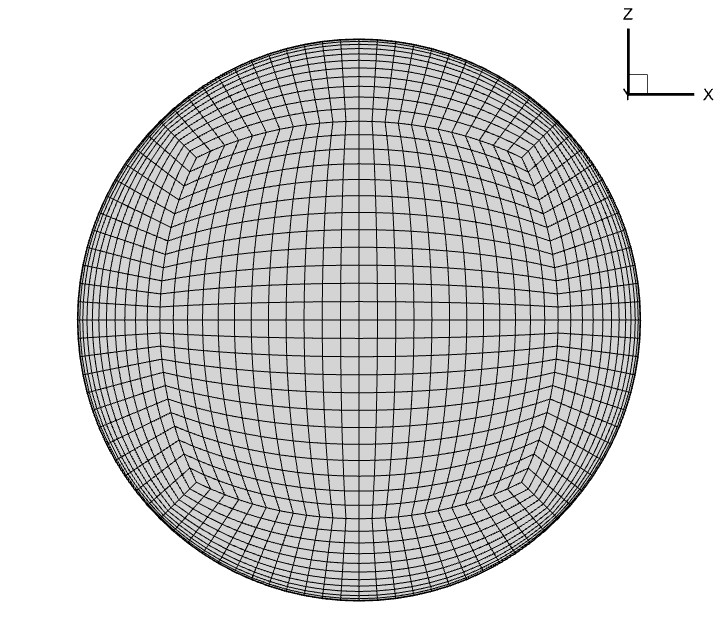}
         \subcaption[a]{}
     \end{subfigure}
     \hfill
     \begin{subfigure}[b]{0.48\textwidth}
         \centering
         \includegraphics[width=\textwidth]{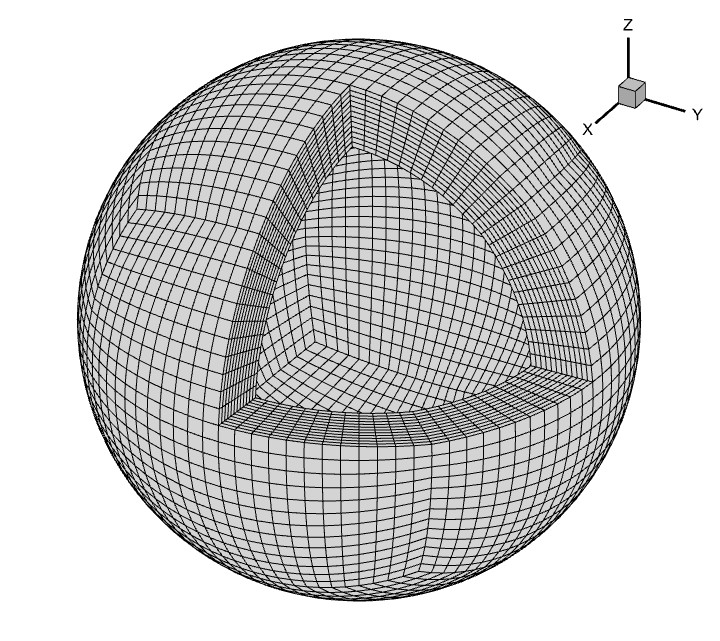}
         \subcaption[a]{}
     \end{subfigure}
     \hfill
        \caption{Cubed Sphere Mesh with 12 Radial Cells and 24 Zonal Cells}
        \label{Grid double}
\end{figure}

The spherical shell is generated by layering a series of surface meshes. CHORUS-MHD uses a cubed-sphere meshing technique, which projects the six faces of a cube onto the spherical surface for each radial increment. In total, there are $N_{r}$ radial increments. Using a cubed-sphere mesh allows for a given face to have angularly equidistant hexahedral elements for all radii. Angularly equidistant elements on each face prevent singularities and allow for better resolution in the polar regions. It is also possible to generate oblate meshes for CHORUS-MHD to run on. 
For one of the six faces, the angles $\alpha=\arctan(\frac{y}{x})$ and $\beta=\arctan(\frac{z}{x})$ are chosen, with $\alpha=\left[-\frac{\pi}{4},\frac{\pi}{4}\right]$ and $\beta=\left[-\frac{\pi}{4},\frac{\pi}{4}\right]$. On each face there are $N_{z}$ by $N_{z}$ zonal elements. An element face at a given radial increment R is defined as the surface enclosed by four arcs. The pairs of arcs enclosing the $i^{th},j^{th}$ element on a single face are described by the angles
\begin{equation}
\left[\alpha_{i},\alpha_{i+1}\right]\times\left[\beta_{i},\beta_{i+1}\right]
\end{equation}
with 
\begin{equation}
\alpha_{i}=-\frac{\pi}{4}+\frac{i}{N_{z}}\frac{\pi}{2} \ \ \ \ \ \ \ \ \ \ \ \ \ \ \ 
\beta_{i}=-\frac{\pi}{4}+\frac{j}{N_{z}}\frac{\pi}{2}
\end{equation}
The total number of elements in the mesh is given by $N_{cells}=6N_{r}N_{z}^{2}$. The number of degrees of freedom in a three-dimensional mesh is then calculated as $N_{DOF}=N_{cells}\times N_{order}^3$, where $N_{order}$ is the order of the spectral difference method used. Figure \ref{Grid double} shows a cubed sphere mesh with $N_r=12$ and $N_z=24$.
\section{Spectral Difference Method}
\begin{figure}[h]
    \centering
    \includegraphics[width=0.7\linewidth]{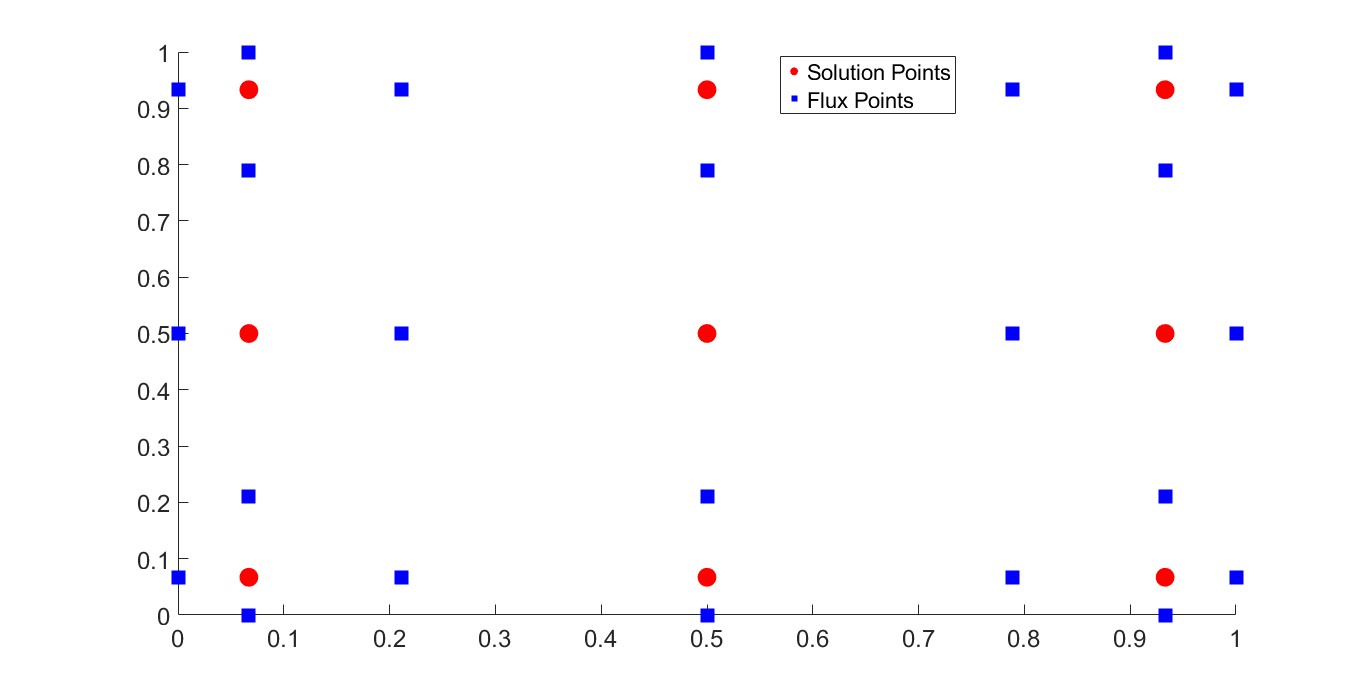}
    \caption{Standard element solution Points (red) and flux Points (blue) for a two dimensional, third order spectral difference method}
    \label{SPsFPs}
\end{figure}

To perform calculations in CHORUS-MHD, all elements are transformed into a standard cubic element in the computational domain. The transformation is described as
\begin{equation}
    \left(x,y,z\right)=\mathbf{P}\left(\xi,\eta,\zeta\right)
\end{equation}
Where $\xi$, $\eta$, and $\zeta$ are the coordinates in the computational domain defined as
\begin{equation}
    \left(\xi,\eta,\zeta\right)\in \left[0,1\right]\times\left[0,1\right]\times\left[0,1\right]
\end{equation}
The transfinite mapping $\textbf{P}\left(\xi,\eta,\zeta\right)$ is created using linear combinations of projectors, including bilinear projectors and a trilinear projector as described in \cite{Chen2023}. To perform computations the governing equations are transformed to the computational domain using a Jacobi matrix for the mapping $\textbf{P}$. The Jacobi matrix is defined as:
\begin{equation}
    J=\frac{\partial(x,y,z)}{\partial(\xi,\eta,\zeta)}=\begin{bmatrix}
        x_\xi \ x_\eta \ x_\zeta \\ y_\xi \ y_\eta \ y_\zeta \\z_\xi \ z_\eta \ z_\zeta
    \end{bmatrix}
\end{equation}
Using the Jacobi matrix, the governing equations in the form of Equation \ref{divergence form} can be written as 
\begin{equation}
    \frac{\partial \tilde{\mathbf{Q}}}{\partial t} + \mathbf{\nabla} \cdot \tilde{\mathbf{F}} = \tilde{\mathbf{M}}
    \label{G.E. comp}
\end{equation}
The solution vector and source vector are transformed as $\tilde{\mathbf{Q}}=|J|\mathbf{Q}$ and $\tilde{\mathbf{M}}=|J|\mathbf{M}$. The flux vector is transformed as
\begin{equation}
\tilde{\mathbf{F}}=|J|J^{-1}\bar{\mathbf{F}}
\end{equation}
Once in the computational domain, for an $N^{th}$ order spectral difference method, a given cubic element has $N^3$ solution points, forming a 3-D grid of solution points. For each iteration, the divergence of the flux vector must be determined, which requires a $(N+1)^3$ grid of flux points in each cubic element to maintain order. A standard 2-D element of flux points and solution points is shown in Figure \ref{SPsFPs} for reference.
Both the solution point and flux point grids are created by repeating 1-D spacings in each coordinate direction. The solution points are chosen to be Chebyshev–Gauss quadrature points. The $n^{th}$ solution point is given as
\begin{equation}
    X_{s,n}=\frac{1}{2}\left[1-cos\left(\frac{2n-1}{2N}\pi\right) \right] \ , \ n=1,2,...N
\end{equation}
The flux points $X_{f}$ are roots of the Legendre polynomials of $N^{th}$ degree with two additional endpoints on the cell boundaries at zero and one. As seen in Figure \ref{SPsFPs}, each string of solution points, in either unit direction, has an associated string of flux points in the same unit direction used for Lagrange interpolation. For a two-dimensional grid, the flux points are split into two families. In the $\xi$ direction the family of points is given as: $(\xi,\eta)=(X_{f,i},X_{s,j})$ for $i=[1,N+1]$ and $j=[1,N]$. Similarly in the $\eta$ direction, the family is described as: $(\xi,\eta)=(X_{s,i},X_{f,j})$ for $i=[1,N]$ and $j=[1,N+1]$.
\\ \indent To find the time derivative of the solution vector at a given solution point, the divergence of the flux must be evaluated at $X_s$, as seen in Equation \ref{G.E. comp}. Each iteration starts with the solutions known at the solution points, $X_s$. The solutions and solution gradient are then extrapolated to their respective families of flux points using one-dimensional Lagrange polynomials. The flux is then calculated using Equation \ref{invF} and Equation \ref{visF}, and a Lagrange polynomial of order $N+1$ is constructed of the flux. The gradient of the flux polynomial is then calculated and evaluated at the solution points. Lagrange polynomials are constructed using Lagrange basis functions at the solution points and the flux points in the $(\xi,\eta,\zeta)$ directions. For the solution points the Lagrange basis function is
\begin{equation}
    h_i(X) = \prod_{s=1,s\neq i}^{N} \left(\frac{X-X_s}{X_i-X_s}\right)
\end{equation}
Similarly for the flux points, 
\begin{equation}
    l_i(X) = \prod_{f=1,f\neq i}^{N+1} \left(\frac{X-X_f}{X_i-X_f}\right)
\end{equation}
To reconstruct solution and flux polynomials in the computational domain, is the tensor product of the one dimensional Lagrange polynomials. For the solutions in the computational domain
\begin{equation}
   \mathbf{\tilde{Q}}(\xi,\eta,\zeta) = \sum_i^N \sum_j^N \sum_k^N \frac{\mathbf{\tilde{Q}}_{i,j,k}}{|J_{i,j,k}|}h_i(\xi) h_j(\eta)h_k(\zeta)
\end{equation}
For the flux in the computational domain the components of the flux vector $\mathbf{\tilde{F}}=\tilde{F}\hat{\xi}+\tilde{G}\hat{\eta}+\tilde{H}\hat{\zeta}$ are given as
\begin{equation}
    \tilde{F}(\xi,\eta,\zeta) = \sum_i^N \sum_j^N \sum_k^N \mathbf{\tilde{F}_{i,j,k}}l_i(\xi) h_j(\eta)h_k(\zeta)
\end{equation}
\begin{equation}
    \tilde{G}(\xi,\eta,\zeta) = \sum_i^N \sum_j^N \sum_k^N \mathbf{\tilde{G}_{i,j,k}}h_i(\xi) l_j(\eta)h_k(\zeta)
\end{equation}
\begin{equation}
    \tilde{H}(\xi,\eta,\zeta) = \sum_i^N \sum_j^N \sum_k^N \mathbf{\tilde{H}_{i,j,k}}h_i(\xi) h_j(\eta)l_k(\zeta)
\end{equation}
To ensure solution polynomials are continuous across element faces, the common solution evaluated at element interfaces is taken as the average of the polynomials from both cells, known as the BR1 scheme \cite{Bassi1997}. A similar procedure is used for the viscous flux across cell interfaces. For inviscid flux, the Rusanov solver \cite{Rusanov1962} is used to ensure a common flux. The cell face averaging processes are performed during the previously described iteration procedure, after the construction of a solution or flux polynomial.
\\ \indent Once the spatial derivatives of the flux polynomials are computed, the time derivative of the solutions can be found. For time stepping, CHORUS MHD uses a five-stage third-order explicit strong stability-preserving Runge-Kutta method [SSPRK(5,3)]. The coefficients for the method are tabulated in Table 1 of Ruuth \cite{Ruuth2006}.

\section{Initial Conditions}
Initially all conserved variables are only a function of radius, and the shell is in hydrostatic balance. Mathematically, $\mathbf{U}=0$, and 
\begin{equation}
    \frac{dp}{dr}=-\rho g
    \label{hydrostatic}
\end{equation}
The classical solutions for static stratification give initial profiles for thermodynamic variables $\rho$,$p$,$T$ are
\begin{equation}
    p=p_{b}\left[1-\frac{\Phi-\Phi_{b}}{C_{p}T_{b}}\right]^\frac{\gamma}{\gamma-1} \ \ \ \  \ \ \ 
    \rho=\rho_{b}\left[1-\frac{\Phi-\Phi_{b}}{C_{p}T_{b}}\right]^\frac{1}{\gamma-1} \ \ \ \ \ \ \
    T=T_{b}\left[1-\frac{\Phi-\Phi_{b}}{C_{p}T_{b}}\right]
    \label{polytrop eqns}
\end{equation}
Where $\Phi$ is the gravitational potential given by $\Phi=-\frac{GM_{\odot}}{r}$ and $T_{b}$ is the temperature at the bottom surface given by
\begin{equation}
T_{b}=\frac{\Phi_t-\Phi_b}{C_p(1-e^{-(\gamma-1) N_\rho})}
\end{equation}
Where $N_\rho$ is the number of density scale heights defined as $N_\rho=\ln\left(\frac{\rho_b}{\rho_t}\right)$. The relations listed in Equation \ref{polytrop eqns} do not satisfy thermal equilibrium. To satisfy an approximate thermal equilibrium, an almost flux-balanced approach is used \cite{Wang2015}. In the initial state, the total heat flux from the shell is constant through each radial layer. At the bottom boundary, the heat flux can be defined by
\begin{equation}
        \mathbf{q}=-\kappa \rho T \mathbf{\mathbf{\nabla}} S - \kappa_r \rho C_p \mathbf{\nabla} T=\frac{L_\odot}{4\pi r^2_b}\mathbf{\hat{r}}       \label{qfluxIC}
\end{equation}
Where $L_\odot$ is the stellar luminosity of the spherical shell, defined as an input. The gradients $\mathbf{\nabla} S$ and $\mathbf{\nabla} T$ simplify to $\frac{dS}{dr}$ and $\frac{dT}{dr}$ respectively. Rearranging Equation \ref{qfluxIC} for the entropy gradient gives 
\begin{equation}
        \frac{dS}{dr}=\frac{1}{\kappa\rho T}\left(\frac{L_\odot}{4\pi r^2_b}+\kappa_r \rho C_p \frac{dT}{dr}\right)
        \label{dSdr}
\end{equation}
Using Equations in \ref{polytrop eqns} and Equation \ref{dSdr}, the target entropy gradient can be solved for using the form
\begin{equation}
        S=S_b+\int^r_{r_b}\frac{dS}{dr}dr
\end{equation}
Once the entropy profile is computed, we can then solve for a new density gradient. Using Equations \ref{eq:entropy} and \ref{hydrostatic}, it can be shown that 
\begin{equation}
        \frac{d\rho}{dr}=-\left(\frac{\rho}{C_p}\frac{dS}{dr}+\frac{g}{\gamma}\rho^{2-\gamma}e^\frac{\gamma S}{C_p}\right)
        \label{densityprofile}
\end{equation}
With the entropy $S$ and its gradient $\frac{dS}{dr}$ already computed, Equation \ref{densityprofile} can be solved numerically for the new density distribution. Once the new density distribution is calculated, the pressure and temperature distributions can also be calculated. The system starts thermally stable, without convection. The constant temperature flux from the bottom boundary of the shell increases the temperature gradient until it becomes critical, resulting in thermal instability and convection.

\section{Problem Formulation and Boundary Conditions}
CHORUS considers a hollow fluid shell with most of the mass inside the shell. A constant heat flux is imposed on the inner surface to induce convection, and the outer surface has a constant temperature. The inner and outer surfaces are assumed to be stress-free and impenetrable. For the magnetic boundary conditions, the surfaces can be defined as perfectly electrically conducting, or a perfectly radial magnetic field can be defined. The constant heat flux is at the bottom surface is given by
\begin{equation}
\mathbf{q}=\frac{L_\odot}{4\pi r^2_b}\mathbf{\hat{r}}
\end{equation}
$L_\odot$ is the stellar luminosity of the spherical shell, specified as an input, and $r_b$ is the radius of the inner surface. Impenetrable boundaries are described by Equation \ref{inpen boundaries}.
\begin{equation}
\mathbf{U} \cdot \mathbf{\hat{r}}=0
\label{inpen boundaries}
\end{equation}
For the stress-free surfaces, the shear stress tensor $\mathbf{\bar\tau}$ is converted to spherical coordinates, and the angular components are set to zero for cells at $r=r_b$ and $r=r_t$. The perfectly conducting boundary conditions are described as $\mathbf{J}=J_r\mathbf{\hat{r}}$ at $r=r_s$ where $r_s$ is the surface to which the boundary conditions are applied. Using $\mathbf{J}=\mathbf{\nabla}\times\mathbf{B}$ leads to the magnetic field conditions that $B_r=\frac{\partial}{\partial r}\left(rB_\theta\right)=\frac{\partial}{\partial r}\left(rB_\phi\right)=0$ at $r=r_s$. In CHORUS-MHD, this is implemented by transforming the Jacobian of the magnetic field into spherical coordinates and setting the appropriate derivatives to zero. For perfectly radial conditions $B_\theta=B_\phi=0$ at $r=r_s$, and to maintain the divergence-free magnetic field, $\frac{\partial}{\partial r}\left(r^2B_r\right)=0$ at $r=r_s$. For perfectly insulating boundaries, $\mathbf{J}=0$ at $r=r_s$ and outside of the spherical shell. To model perfectly insulating boundary conditions, now that $\mathbf{\nabla} \times\mathbf{B}=0$, the magnetic field can now be represented as a scalar potential, and the magnetic field inside of the shell can be matched to the magnetic field outside of the shell. For codes using spherical harmonics, the perfectly insulated boundary conditions are rather easy to apply by mapping the magnetic field to the scalar magnetic field potential. For fully compressible codes using a method other than spherical harmonics, perfectly insulating boundary conditions would be much more cumbersome. Mapping the magnetic field to the potential magnetic field requires solving the Laplace equation numerically for the scalar magnetic field at the given boundary of the shell. Currently, perfectly insulating boundary conditions are not implemented in CHORUS-MHD. In this study, benchmark tests were run with perfectly radial boundary conditions, which allow surface current on the shell.

\section{Benchmark Problem Definition}
Two benchmark problems are presented for CHORUS-MHD. The first benchmark test is an extension of a hydrodynamic model of the Sun presented by Wang et al. \cite{Wang2015} to an MHD model. The key parameters defining the solar benchmark test are shown in Table \ref{SolarParameters}. The second benchmark test increases the density scale height from $N_\rho=3$ to $N_\rho=4$. Both simulations use a sixth-order spectral difference method.
\begin{table}[h]
    \centering
    \caption{Solar Benchmark Parameters}
    \begin{tabular}{|l|l|l|}
        \hline
        \multicolumn{3}{|c|}{\textbf{Physical Input Parameters:}} \\ \hline
        $r_t = 6.61 \times 10^{10} \, \text{cm}$ & $r_b = 4.87 \times 10^{10} \, \text{cm}$ & $M_\odot = 1.98891 \times 10^{33} \, \text{g}$  \\ \hline
        $\mathbf{\Omega} = 8.1 \times 10^{-5} \, \text{s}^{-1}$ & $\rho = 0.21 \, \text{g cm}^{-3}$ & $G = 6.67 \times 10^{-8} \, \text{g}^{-1} \, \text{cm}^3 \, \text{s}^{-2}$  \\ \hline
        \multicolumn{3}{|c|}{\textbf{Fluid Properties:}} \\ \hline
        $R = 1.4 \times 10^8 \, \text{erg} \text{ g}^{-1} \, \text{K}^{-1}$ & $\kappa = 6.0 \times 10^{13} \, \text{cm}^2 \, \text{s}^{-1}$ & $\nu = 6.0 \times 10^{13} \, \text{cm}^2 \, \text{s}^{-1}$  \\ \hline
        $\gamma = \frac{5}{3}$ & $\eta=1.2\times10^{13} \ cm^2 \ s^{-1}$& \\ \hline
        \multicolumn{3}{|c|}{\textbf{Thermodynamic Properties:}} \\ \hline
        $L_\odot = 3.846 \times 10^{36} \, \text{erg s}^{-1}$ & $C_p = 3.5 \times 10^8 \, \text{erg g}^{-1} \, \text{K}^{-1}$ & $N_\rho=3$ \\ \hline
    \end{tabular}
    \label{SolarParameters}
\end{table}

The radiative flux of the solar benchmark is given by $\kappa_r=\lambda\left(c_0+c_1\omega+c_2\omega^2\right)$ with $c_0=1.5600975\times10^8$, $c_1=-4.5631718\times10^7$, $c_2=3.3370368\times10^6$, and $\omega=r\times10^{-10}$.  For the solar benchmark, the lower surface is assumed to be perfectly conducting as the magnetic flux is typically very small. For the upper surface, the magnetic field is assumed to be perfectly radial, following observations of the Sun’s magnetic field near the surface.

\section{Results}
For both simulations, the globally averaged kinetic energy, globally averaged magnetic field, meridional circulation, differential rotation, poloidal magnetic energy, and toroidal magnetic energy are calculated. The averaged kinetic energy and magnetic field energy are given as
\begin{equation}
    KE=\frac{1}{{V}} \int_V\frac{1}{2}\rho\mathbf{U}\cdot\mathbf{U}dV
\end{equation}

\begin{equation}
    E_B=\frac{1}{{V}} \int_V\frac{1}{2}\mathbf{B}\cdot\mathbf{B}dV
\end{equation}
The differential rotation presented is simply the longitudinal-time average of $V_\phi$, and the meridional flow is the combination of the radial velocity $V_r$ and the polar angle velocity $V_\theta$ as $V_{mer}=\sqrt{V_r^2+V_\theta^2}$, longitudinally time averaged.
The poloidal and toroidal contributions to the magnetic field energy are given as 
\begin{equation}
    E_{Bpol}=\frac{B_r^2+B_{\theta}^2}{8\pi} \ \ \ \ \ \ \ \ \ \ \ \ \ \ \ \ \ \ \ \ E_{Btor}=\frac{B_{\phi}^2}{8\pi}
\end{equation}

\subsection{First Solar Benchmark}
\begin{figure}
    \centering
    \includegraphics[width=0.4\linewidth]{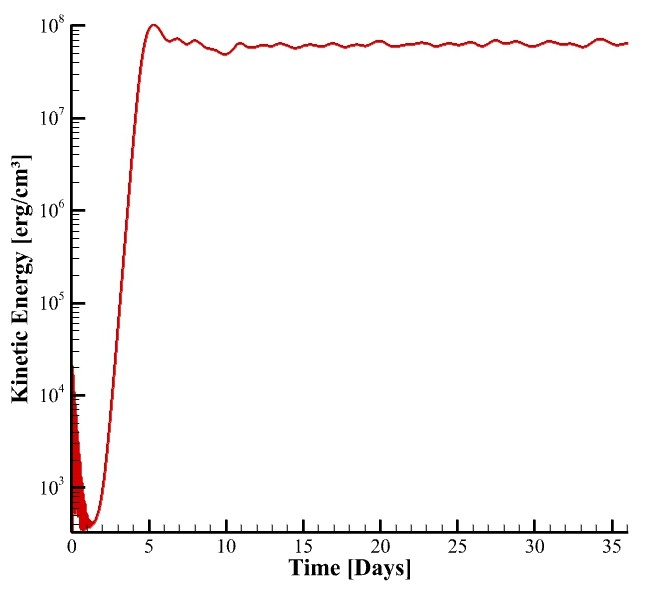}
    \caption{Globally Average Kinetic Energy for First Benchmark Test}
    \label{KE3}
\end{figure}

\begin{figure}

\begin{subfigure}[a]{0.45\textwidth}
         \centering
         \includegraphics[width=\textwidth]{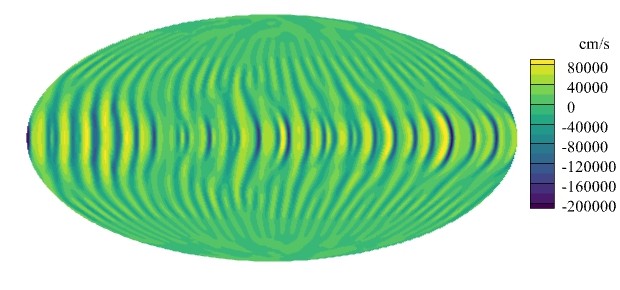}
         \subcaption[a]{$V_r$ Mollweide Projection at R=0.95}
     \end{subfigure}
     \hfill
    \begin{subfigure}[a]{0.45\textwidth}
         \centering
         \includegraphics[width=\textwidth]{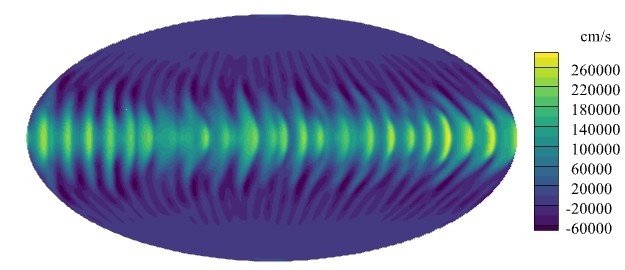}
         \subcaption[a]{$V_\phi$ Mollweide Projection at R=0.95}
     \end{subfigure}
     \hfill
    \begin{subfigure}[c]{0.45\textwidth}
         \centering
         \includegraphics[width=\textwidth]{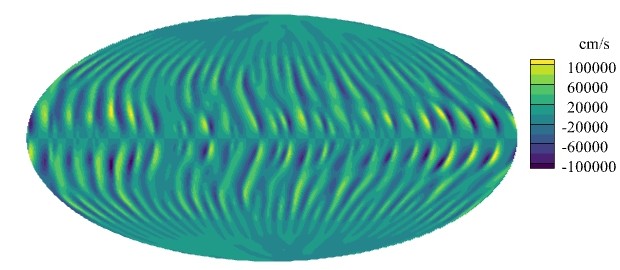}
         \subcaption[c]{$V_\theta$ Mollweide Projection at R=0.95}
     \end{subfigure}
     \hfill
    \begin{subfigure}[d]{0.45\textwidth}
         \centering
         \includegraphics[width=\textwidth]{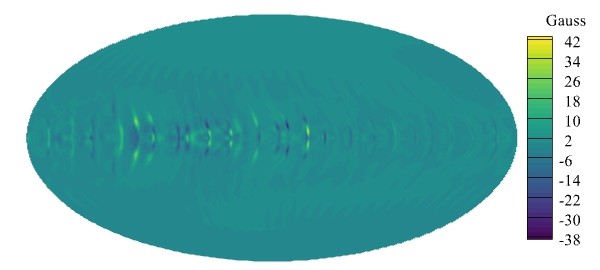}
         \subcaption[c]{$B_r$ Mollweide Projection at R=0.95}
     \end{subfigure}
     \hfill
    \begin{subfigure}[d]{0.45\textwidth}   
     \centering
\includegraphics[width=\textwidth]{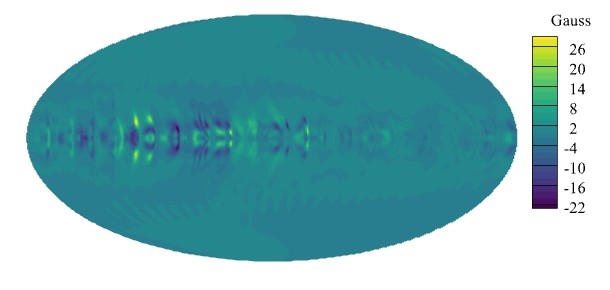}
         \subcaption[e]{$B_\phi$ Mollweide Projection at R=0.95}
     \end{subfigure}
     \hfill
    \begin{subfigure}[g]{0.45\textwidth}
         \centering
         \includegraphics[width=\textwidth]{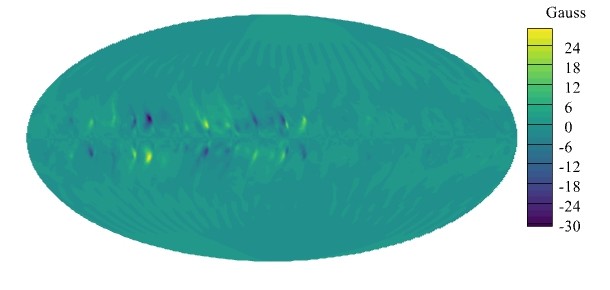}
         \subcaption[f]{$B_\theta$ Mollweide Projection at R=0.95}
     \end{subfigure}
        \caption{Mollweide Projections of Velocity Fields and Magnetic Fields for Benchmark 1}
        \label{SolarMollweide}
\end{figure}

The first benchmark was run using Clarkson University's GPU partition on ACRES. The simulation was run using 3 GPUs, with a mesh using 28 radial and 26 zonal cells, for a total of 36 simulated days. CHORUS-MHD produces a lower steady-state kinetic energy density of approximately $6.47\times10^7 \  erg \ cm^{-3}$ in comparison to the hydrodynamic benchmark, as seen in Figure \ref{KE3}. A lower steady state energy is to be expected in a full MHD simulation due to the addition of a magnetic field. Magnetic forces restrict fluid flow perpendicular to the magnetic field lines, and energy is constantly being converted between the magnetic field and the velocity field.
\begin{figure}
 \begin{subfigure}[a]{0.24\textwidth}
         \centering
         \includegraphics[width=\textwidth]{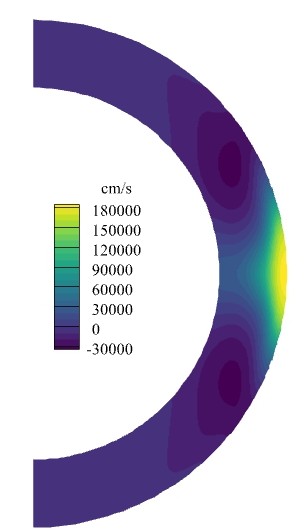}
         \subcaption[a]{Differential Rotation}
     \end{subfigure}
     \hfill
    \begin{subfigure}[a]{0.225\textwidth}
         \centering
         \includegraphics[width=\textwidth]{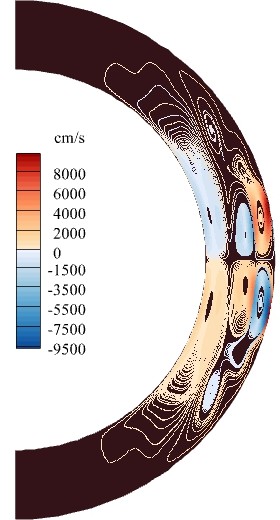}
         \subcaption[a]{Meridional Circulation}
     \end{subfigure}
     \hfill
    \begin{subfigure}[c]{0.24\textwidth}
         \centering
         \includegraphics[width=\textwidth]{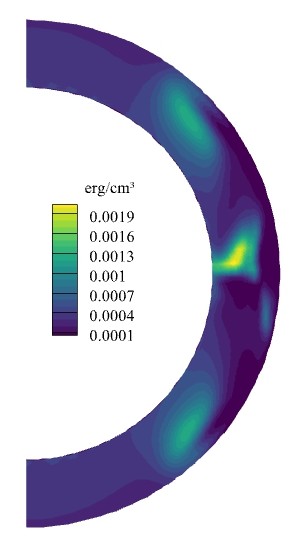}
         \subcaption[c]{Poloidal Magnetic Energy}
     \end{subfigure}
     \hfill
    \begin{subfigure}[d]{0.24\textwidth}
         \centering
         \includegraphics[width=\textwidth]{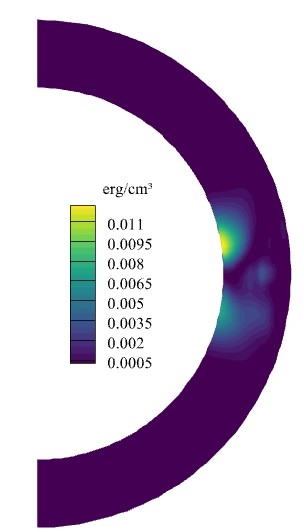}
         \subcaption[c]{Toroidal Magnetic Energy}
     \end{subfigure}
     \hfill
        \caption{Longitudinal Time-Averaged Velocity Fields and Magnetic Fields for Benchmark 1}
        \label{SolarLonAvg}
\end{figure}
Figure \ref{SolarLonAvg} shows longitudinally time-averaged plots of the velocity field and magnetic field. Figure \ref{SolarMollweide} shows the mollweide plots of the velocity field and magnetic field. As seen in the longitudinal-time average contours, the MHD simulation presents strong differential rotation and meridional circulation, similar to the hydrodynamic benchmark. The meridional circulation predicts a flow pattern similar to that of Wang et al. \cite{Wang2015}, with a near-surface circulation, followed by two deeper circulations in the opposite direction. This meridional flow pattern is reflected in both the upper and lower hemispheres. In the Mollweide projections, MHD CHORUS predicts banana cells similar to that of the hydrodynamic benchmark, with many upflow lanes separated by stronger downflow lanes, however with a reduced flow magnitude. The lower kinetic energy and velocity profiles are evident that the magnetic field is key to reducing typical over predictions of fluid velocity. The magnetic field is mostly concentrated near the equator for all three magnetic field components. The concentration of the magnetic field near the equator is likely due to the increased flow speed near the equator, seen in both the differential rotation and the meridional flow. In the shown Mollweide projection, note that all magnetic-field components are concentrated at a similar location in the left hemisphere. In the velocity fields, most notably the radial and azimuthal fields, there is a dampened zone in the same region where the magnetic field is peaking. This is a similar behavior seen in solar activity like sunspots, where the flow is restricted due to magnetic field lines, and the area's luminosity is lowered due to the reduced supply of plasma.

\subsection{Second Solar Benchmark with Increased Density Scale Height}
\begin{figure}
     \centering
     \begin{subfigure}[a]{0.4\textwidth}
         \centering
         \includegraphics[width=\textwidth]{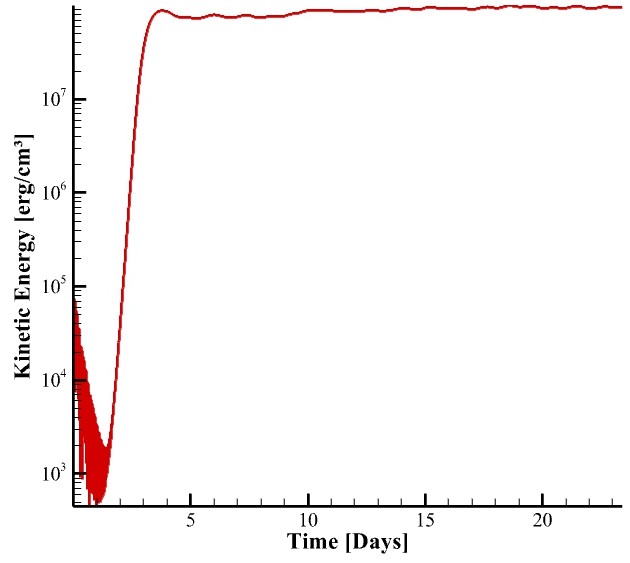}
     \end{subfigure}
     \hfill
     \centering
     \begin{subfigure}[a]{0.4\textwidth}
         \centering
         \includegraphics[width=\textwidth]{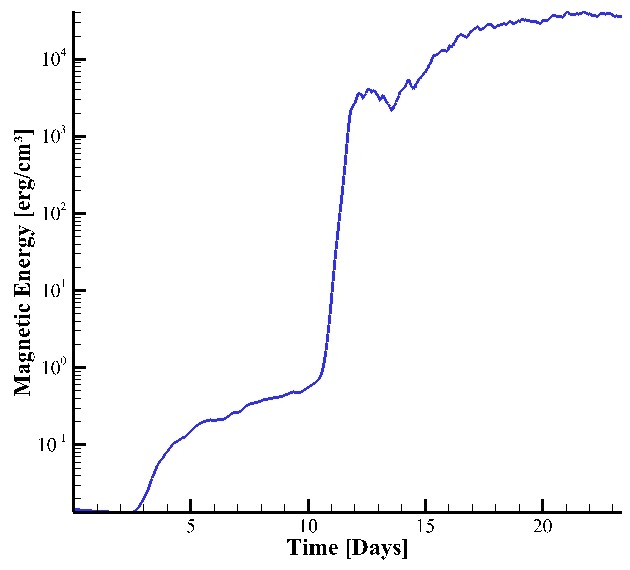}
     \end{subfigure}
     \hfill
     \caption{Globally Average Kinetic Energy (a) and Globally Averaged Magnetic Energy (b)}
        \label{SolarKEplot}
\end{figure}

\begin{figure}
\centering
    \hfill
    \begin{subfigure}[a]{0.21\textwidth}
         \centering
         \includegraphics[width=\textwidth]{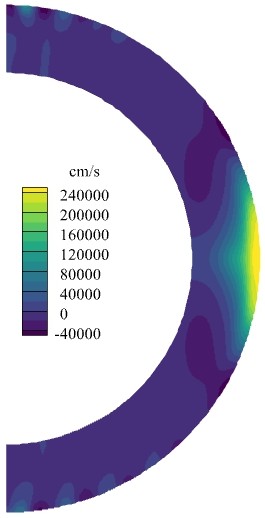}
         \subcaption[a]{Differential Rotation}
     \end{subfigure}
     \hfill
    \begin{subfigure}[a]{0.225\textwidth}
         \centering
         \includegraphics[width=\textwidth]{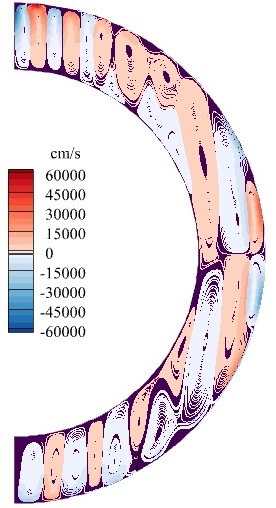}
         \subcaption[a]{Meridional Circulation}
     \end{subfigure}
     \hfill
    \begin{subfigure}[c]{0.23\textwidth}
         \centering
         \includegraphics[width=\textwidth]{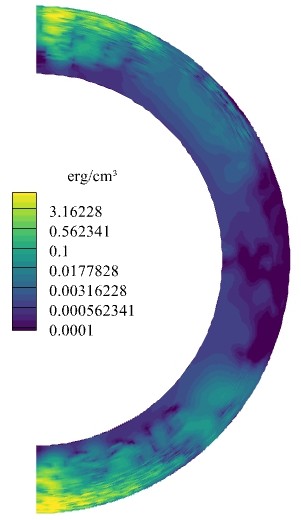}
         \subcaption[c]{Poloidal Magnetic Energy}
     \end{subfigure}
     \hfill
    \begin{subfigure}[d]{0.23\textwidth}
         \centering
         \includegraphics[width=\textwidth]{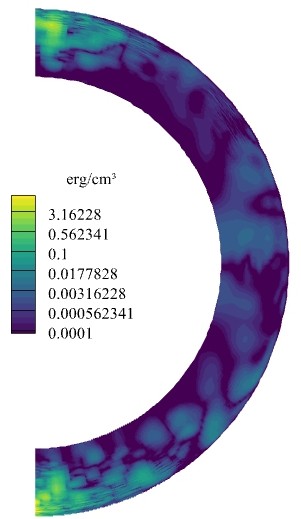}
         \subcaption[c]{Toroidal Magnetic Energy}
     \end{subfigure}
     \hfill
        \caption{Longitudinally Time-Averaged Velocity Fields and Magnetic Fields for Benchmark 2}
        \label{SolarLonAvgRho4}
\end{figure}

The second benchmark was run using NASA's Pleiades GPU partition. The simulation was run using 16 GPUS, with a mesh using 32 radial and 34 zonal cells, for a total of simulated 24 days. Figure \ref{SolarKEplot} shows the globally averaged kinetic energy (a) and the globally averaged magnetic energy (b) vs time. Figure \ref{SolarLonAvgRho4} shows longitudinally time-averaged plots of the velocity field and magnetic field for the second solar benchmark. Magnetic field plots are plotted on an exponential basis because of the large contrast in magnitudes between the equatorial and polar regions.
The kinetic energy density of the second solar benchmark saturates at approximately $9.57\times10^7 \  erg \ cm^{-3}$. The higher kinetic energy of the second benchmark is probably due to the increase of the density scale height without reducing the luminosity or rotational speed of the simulation. From the longitudinally time-averaged plots of the second benchmark, it can be seen that a pure increase in density scale height has large effects on both the velocity and magnetic fields. First, there is a large increase in the magnitude of both differential flow and meridional circulation, with the meridional circulation having a much larger change. The meridional flow pattern in the second solar benchmark is also significantly different from the first benchmark. In the second case, the near-equator meridional flow switches directions on the innermost circulation, and alternating convective cells can be seen to extend all the way up to the poles with strong magnitudes. From the poloidal and toroidal magnetic energy contours, the polar region is much more active, probably corresponding to the strong convective cells near the poles.
\section{Concluding Remarks}
The CHORUS++ code presented in \cite{Chen2023} is now being further developed successfully to solve 3D MHD equations. CHORUS-MHD is also GPU accelerated in this study before it is applied to study two solar dynamo benchmark problems with different density scale heights. One benchmark problem involves an evident exchange of poloidal and toroidal magnetic energies near the equatorial region. The other benchmark problem captures two magnetically active poles that closely correlate with active convective cells in the polar regions. With both solar benchmark tests completed, the next step in our investigation is to perform further parametric studies of the case with increased density scale height. The next simulation will reduce the model luminosity and rotation rate to better control velocity and magnetic fields, with an expectation to achieve a solar-like meridional-flow profile and cyclic magnetic fields at a higher density scale height.

\section*{Acknowledgments}

The research work in this paper has been financially supported by a National Science Foundation (NSF) award
(No. 2310372) monitored by Dr. Lisa Winter and an Air Force Office of Scientific Research (AFOSR) grant (award No. FA9550-23-1-0596) monitored by Dr. Fariba Fahroo. The computational resources for this work were partially supported by the NASA grant 80NSSC20K0602. Our computations were performed on the GPU nodes of Clarkson's ACRES cluster and NASA's Pleiades Supercomputer. CHORUS-MHD was GPU accelerated recently thanks to the M.S. thesis work by Russell Hankey.

\bibliography{aiaa_CHORUS_MHD}

\end{document}